# Mudslide: A Spatially Anchored Census of Student Confusion for Online Lecture Videos

**Elena L. Glassman**[1,2]    **Juho Kim**[1,2]    **Andrés Monroy-Hernández**[1]    **Meredith Ringel Morris**[1]

[1]Microsoft Research, Redmond, WA, USA
[2]MIT CSAIL, Cambridge, MA, USA
{juhokim, elg}@mit.edu         {andresmh, merrie}@microsoft.com

**ABSTRACT**

Educators have developed an effective technique to get feedback after in-person lectures, called "muddy cards." Students are given time to reflect and write the "muddiest" (least clear) point on an index card, to hand in as they leave class. This practice of assigning end-of-lecture reflection tasks to generate explicit student feedback is well suited for adaptation to the challenge of supporting feedback in online video lectures. We describe the design and evaluation of Mudslide, a prototype system that translates the practice of muddy cards into the realm of online lecture videos. Based on an in-lab study of students and teachers, we find that spatially contextualizing students' muddy point feedback with respect to particular lecture slides is advantageous to both students and teachers. We also reflect on further opportunities for enhancing this feedback method based on teachers' and students' experiences with our prototype.

**ACM Classification Keywords**
H.5.m. Information interfaces and presentation: Misc.

**INTRODUCTION**

Teachers are finding "flipped classrooms" increasingly attractive and feasible [12]. In a flipped classroom, students watch lectures outside of class and do more active learning activities within class. Some teachers flip their classrooms using educational videos from services like Khan Academy [khanacademy.org], Coursera [coursera.org], or edX [edx.org], while others create their own video lectures using tools like Camtasia [techsmith.com] or Office Mix [officemix.com]. One drawback of online lectures is that teachers no longer have access to many traditional signals of student confusion (such as facial expressions or raised hands).

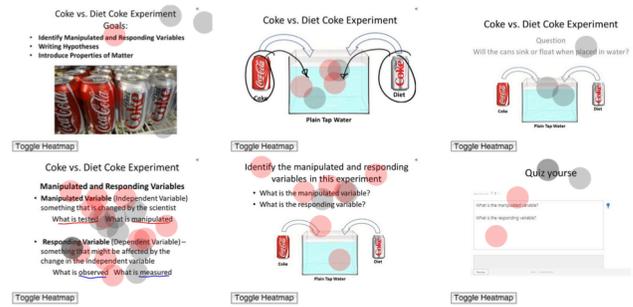

**Figure 1. Mudslide Teacher Interface:** slides presented in the lecture video, with muddy points generated by students that can be clicked to view descriptions of confusion. Red muddy points indicate that the author found the lecture confusing. Gray muddy points are created by those who rated the lecture as not confusing overall.

Educators have developed an effective technique to get feedback about in-person lectures, called "muddy cards" [18] or "One-Minute Papers" [11,22]. These are index cards that teachers give to students at the end of a lecture. Students are given time to reflect and write the "muddiest" (least clear) point on the index card to hand in as they leave. The teacher then adjusts the next day's activities based on students' common confusion points. The aggregation of the responses in the index cards is often a time-consuming process [11,18].

This practice of assigning end-of-lecture reflection tasks to generate explicit student feedback seems well suited for adaptation to the challenge of supporting feedback on online video lectures. In this paper, we describe the design and evaluation of Mudslide, a prototype system that translates the practice of "muddy cards" into the realm of online lecture videos. Mudslide provides timely feedback where real-time feedback, i.e., students' raised hands, is absent. Figure 1 shows Mudslide's teacher interface.

After discussing related work, we describe how we adapt the workflow for soliciting and summarizing muddy points to online lecture technologies, and we present Mudslide, our instantiation of this concept. We then describe a three-stage evaluation of Mudslide, in which students used the tool to give feedback on lecture videos and teachers viewed and reflected on this feedback. Our findings indicate that spatially contextualizing students' muddy point feedback with respect to particular lecture slides is advantageous to both students and teachers; Mudslide's workflow demonstrates a simple yet effective method for incorporating new types of student feedback into online educational experiences.

## RELATED WORK

We first discuss the traditional Muddy Card educational method that inspired our system design. We then discuss literature on the role of confusion in learning. Finally, we reflect on status quo techniques for gathering student feedback from online lectures.

### Muddy Cards

The Muddy Card workflow is an end-of-lecture reflection method for students that has been deployed in a variety of traditional educational settings, from middle school math classrooms to Harvard Statistics [18] and MIT Aero/Astro lecture halls [10]. The Muddy Card workflow is simple: (1) give a lecture, (2) ask students to write what confused them most on an index card, and (3) collect and summarize the responses for further action.

Responses are written at the end of lecture so that by engaging with the material through reflection, students experience benefits like increased retention of the material [2,9]. The Muddy Card workflow turns that engagement into a signal for teachers to learn from and act on, such as by revising their own lectures for future class sections and/or addressing common points of confusion at the start of the next lecture.

While the workflow is simple, the exact prompt used has an impact on the quality of responses. Mosteller [18] observed that asking, *"What do you want to know more about?"* produced answers in his classroom that often just referred back to the key points of the lecture. However, asking a variant of, *"What was the muddiest point in the lecture?"* promoted specific, relevant, and concrete responses describing what students wanted to know more about.

Muddy cards can take a few minutes of students' time at the end of lecture; the demands on teachers' time, however, can be somewhat larger. For instance, after every lecture, Professor Harwood spent 30 minutes or less reviewing Muddy Cards from his ~250-person chemistry class [11]. Professor Mosteller spent 30 to 45 minutes summarizing the most common responses from his ~50 person class, which he then handed back to students, with clarifications [18].

### Beneficial Confusion

Confusion is part of the learning process. Craig et al. [6] observed a positive correlation between confusion and learning. This supports Piaget's theory that cognitive disequilibrium, experienced as confusion, may trigger learning: the creation or restructuring of knowledge schema [13]. However, D'Mello et al. [8] notes that, to be beneficial, confusion must be appropriately resolved. Mudslide aims to expose students' confusion within online lecture videos, so teachers can appropriately resolve it.

### Gathering Feedback on Online Lecture Videos

When students asynchronously view lecture materials online, gathering feedback about whether the lecture was understood or how to improve it for future viewers is not straightforward. Explicit polling is one option; applications like InstFeedback [instfeedback.com] could be used as a virtual Muddy Card. However, the resulting answers are shown as a simple list. This would not address the time-consuming nature of reading and summarizing muddy points in the existing Muddy Card workflow.

Alternatively, teachers can observe online students' confusion by monitoring discussion forums. Discussion forums are a standard feature of many online lecture platforms (e.g., Coursera, Udacity). Researchers have also experimented with novel discussion forum features and formats for educational purposes, such as Nota Bene [24] and WebAnn [4], which are in-place collaborative annotation tools that create a forum within the margins of assigned class reading material. While useful, forums do not necessarily serve as a census of confusion by which teachers can prioritize the development of additional clarifying material. Even with a mechanism for "upvoting" others' questions, it is not clear how many passive student "lurkers" benefit from a teacher answering a particular question. In addition, the bar for participation in forums can be high: students may need to read multiple threads before discovering that their question has not already been asked and answered, and not all feel comfortable posting. This may be reflected in the low forum participation rates observed in massively open online courses (MOOCs): in edX's first offering of "Circuits and Electronics," only 3% of enrolled students participated in the discussion forum [3].

Students' interactions with educational videos can also be an implicit source of feedback. With VidWiki [7], students spatially and temporally annotate lecture videos to correct errors and translate or clarify text for themselves and others. LectureScape [14] converts the log files of students' interactions with lecture videos in MOOCs into a heatmap of viewing activity along the video timeline. Teachers may be able to infer trouble spots, but the explicit reasons for emergent patterns, such as rewatching, are not known.

Voyant [23] and CrowdCrit [17] demonstrate that crowds, given appropriate structure, can generate feedback for content authors. Non-expert or remote crowds created spatially anchored structured feedback specifically for visual designers. However, the feedback-providers were not students; their role is evaluative rather than reflective. Our system and evaluation focus on how spatially anchored, reflective student feedback can be produced and interpreted in the context of online educational lecture videos.

## MUDSLIDE

Mudslide is a prototype system that incorporates ideas from the Muddy Card technique into a feedback system for online lecture videos. The overall system design is minimalistic and general; to use Mudslide, a teacher simply supplies a folder of lecture slide images used in their video (video key frames could be used as an alternative in scenarios where a teacher does not use lecture slides). Mudslide uses this data to create an online, interactive slide gallery, with separate student and teacher interfaces accessible at different URLs. The teacher sends students the student URL for a lecture's Mudslide gallery to visit after viewing an online lecture (for a more seamless experience, the URL could be incorporated into the lecture itself as a link at the end of the video or within a hosting platform's forum area). Mudslide is implemented with the Django web framework, using Python 2.7, HTML5, CSS3, JavaScript, and the d3.js library.

**Design Goals**

In creating Mudslide, we had four design goals, inspired by our review of the literature and by a formative three-hour workshop at a local school with nine teachers (for grades 6 – 12) and three technical staff in which teachers discussed their current practices and concerns with respect to flipping their classrooms while making lecture videos with Office Mix [officemix.com].

- **G1:** Encourage students to reflect on the entire lecture.
- **G2:** Encourage students to provide specific feedback (not just "this lecture was confusing").
- **G3:** Provide a fast and intuitive way for students to give feedback.
- **G4:** Allow teachers to quickly interpret student feedback, even with large class sizes or multiple classrooms.

Toward fulfilling these goals, we modify the freeform responses from the traditional Muddy Card workflow into a two-step process for students at the end of a lecture: (1) indicate the region on a particular slide that represents the muddiest point and (2) provide a typed explanation of why that point was confusing. Since students do not see others' annotations before making their own, their independent judgments may represent a more accurate census of post-lecture confusion than forums [20].

By preserving this reflection as an end-of-lecture activity, we fulfill design goal **G1**: to support students' pedagogically valuable reflection on the lecture as a whole, and their comprehension of each component. Submitted muddy points represent *unresolved* confusion at the end of lecture. Showing the set of lecture slides as the interface to allow students to specify their feedback further supports this reflection by providing a visual reminder of the entire lecture's contents.

Constraining students to generate a spatial anchor for each textual muddy point description is intended to fulfill design goal **G2**: helping students generate specific, concrete comments and questions. It is also a feature intended to fulfill design goal **G3**: to give students a fast and intuitive mechanism for describing muddy points. Rather than explaining both where and why they were confused, they can simply point to where they were confused and then use text to explain why.

The spatial anchors for muddy points are the key to creating a visual summary for teachers, intended to fulfill design goal **G4**: reducing the muddy-point processing burden on teachers. Rather than reading through a list of purely textual muddy points, teachers can first glance at the distribution of spatially anchored muddy points across their lecture slides to assess trouble spots quickly. Many teachers in our workshop teach the same material to multiple classrooms over the course of a day, and could send a single lecture video to many more students than fit in a single classroom.

**Student Interface**

After watching an online lecture and then clicking the Mudslide URL the teacher provided, students see a gallery of thumbnails of the lecture's slides, laid out chronologically. Atop the gallery are instructions stating, "Double-click on *exactly where* this lecture is most confusing. *The exact location of your click will be shown to the teacher.* You may double-click on multiple confusing points, if you wish." Double-clicking anywhere on a slide's thumbnail places a small, semi-translucent circle at that point (Figure 2). After placing a circle, a dialogue box appears (Figure 3) asking the student to explain why that point was unclear. If the student clicks "Cancel" or fails to provide an explanation, the circle is removed. A separate button, beneath the thumbnail gallery, can clear all of that student's muddy points.

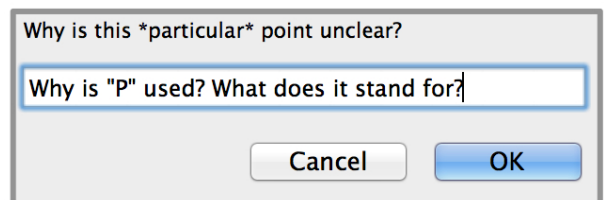

Figure 2. In this example, a student has clicked on the region of a slide stating "if P," creating a red muddy point indicator.

Figure 3. After spatially selecting a muddy point (Figure 2), the student is prompted to provide a specific explanation of why she found that section of the lesson confusing.

Before submitting, students must rate the lecture by choosing one of four choices to complete the following statement: "This lecture was: [extremely confusing | moderately confusing | slightly confusing | not confusing]." After rating the lecture's overall clarity and marking up the

slides with one or more muddy point(s), the student clicks the "Submit to Teacher" button below the last lecture slide.

**Teacher Interface**

The teacher's Mudslide URL directs her to a page that aggregates and summarizes all of the students' feedback. As shown in Figure 1, the defining feature of the teacher interface is an interactive heatmap formed by overlaying the accumulated semi-transparent muddy points on the gallery of lecture slide thumbnails. To save teachers time, the slide with the largest number of muddy points is featured, with its heatmap, at the top of the gallery.

The teacher can click on any spot on the heatmap; clicks within the boundary of one or more circles trigger a pop-up displaying the text of the corresponding muddy point descriptions. A button allows the teacher to toggle the heatmap on and off in case the circles overlap enough to block the text, figure, or equation underneath. To see all comments for a slide at once, the teacher can click a button that "flips" the slide over to reveal a scrollable list of all muddy comments associated with that slide.

By visually differentiating comments by the confusion level of their creator, teachers can investigate trouble spots in their lecture as a function of both the volume of comments and the severity of confusion. Mudslide can differentially color-code points in the heatmap based on students' holistic confusion rating–for example, in Figure 1, points submitted by students who rated the lecture overall as "not confusing" are shown in gray, rather than red. Beneath the gallery, we list the raw muddy point descriptions, ordered by slide, as well as a Word Tree [21] and histogram representation of the same textual data.

**EVALUATION**

To measure the impact of augmenting online lecture videos with muddy point feedback, we ran a multi-stage, lab-based user study.

**Comparison Interface**

As a baseline for comparison, we implemented a research prototype implementing the original Muddy Card workflow. Since the original workflow has no spatial anchors for muddy points, we replaced the gallery of slide thumbnails with a simple text input box in the baseline student interface. The text box is a simple digital metaphor for a physical index card collected in class. We kept the look and feel of the teacher interfaces consistent. Both teacher interfaces had a list of raw comments and summary text representations (histograms and Word Trees); only Mudslide had an interactive heatmap. Since the baseline raw comments have no associated slide, they are ordered by how confusing the student rated the lecture.

We included this baseline interface to allow us to evaluate how spatial anchoring affects students' perceptions of the feedback process, the nature of the feedback provided, and teachers' ability to interpret the feedback. In the analysis that follows, a "muddy card" is the collection of muddy points described by a student in either interface.

**Stage 1. Teachers Create Lecture Videos**

We recruited 19 teachers (13 female) from a U.S. metropolitan region, 14 from public schools, and five from private schools. The teachers (T1-T19) all taught at the middle or high school level (U.S. grades 6–12), and taught a variety of subjects, including math, science, English, and history. Ten of the 19 participating teachers indicated that they had made instructional videos before, using software such as Camtasia, Adobe Premiere, and iMovie. Seven indicated that they had flipped their classrooms before. In a 5-point Likert scale question (1-not familiar, 5-very familiar), most teachers said that they were familiar with PowerPoint (mean=3.8, $\sigma$=0.6). Five had prior experience with Office Mix. The teachers were offered a gratuity to come to our lab for a three-hour session. The teachers were told that they would be preparing video lectures, and each brought PowerPoint files of two of their existing classroom lectures to serve as seed material for their videos.

We provided each teacher with a tablet computer, stylus, webcam, and microphone to use during the study. After completing a questionnaire about personal demographics and any relevant experiences with video lectures, we conducted a tutorial on how to transform their PowerPoint slides into video lectures using Office Mix, a free PowerPoint add-in [officemix.com]. The teachers used the remaining session time to create one or two video lectures based on the slides they brought with them; we instructed them to limit the length of each video to about five minutes (to bound student viewing time in stage 2 of the study).

The teachers made a total of 36 videos. For each video, we asked them to specify the target grade level(s), and to identify what they thought the most confusing point of the lecture might be for students. From the 36 videos, we chose one video authored by each teacher (19 total) for the next stage of the study, based on length, quality, and clarity.

**Stage 2. Students Generate Muddy Points**

We invited 25 middle and high school students (15 female) from a U.S. metropolitan area to our lab (mean age=14.7, max=17, min=13). Each session lasted one hour, and the participants were offered a gratuity for their time. Due to practical constraints, these students were not the members of the 19 teachers' regular classes.

We matched each student (S1-S25) to two of the lecture videos produced in the first stage of the study based on the students' grade level. We also attempted to balance the matching of videos to students so that each video had a similar number of students view it (25 students x 2 videos viewed = 50 viewings divided by 19 available videos = a target of showing each video to 2 – 3 of the students). Students were provided with a tablet computer, mouse, and stylus. We first gave them a questionnaire about their familiarity with online lecture videos, followed by a tutorial on how to view videos using the Office Mix video player. We then demonstrated the features of Mudslide's student interface on a sample set of slides.

Students then completed the following routine twice: (1) watch the assigned lecture video, (2) submit muddy point(s) with the assigned student interface, and (3) complete a survey about their experience. Students were permitted to refer back to the video when composing their muddiest points, if desired. We encouraged students to pick the "least clear" point(s) if they felt that the lecture was clear overall. This stage followed a counter-balanced, within-subjects experimental design. Each student used both the baseline and Mudslide student interfaces once over the course of the session. After the first video, half the students saw the Mudslide student interface, and half saw the baseline.

At the end of the session, the students each filled out a final survey, in which they explicitly compared the baseline and Mudslide student interfaces. The Mudslide student interface was identified by its affordance for pointing and clicking on slides, while the baseline student interface was identified by its text box. Neither interface was ever referred to by a name. Twenty-one of the students completed the study with spare time remaining; we showed these students the Mudslide teacher interface to get formative opinions on whether viewing other students' confusion points was potentially valuable.

The teacher interface is designed to aggregate muddy points from a large or multiple classroom-sized viewer pool. Since time constraints made it impractical to generate a classroom-sized viewer pool of students for each of the teachers' 19 videos, we also hired workers on Amazon's Mechanical Turk service ("Turkers") to view each video and create muddy cards using either the Mudslide or baseline interface. The HITs were completed over ten days with an average hourly wage of $9.04. Based on demographic information requested in the HIT, 34% of muddy cards were composed by Turkers under 25, 37% by women, and 62% by those with at least a college degree.

### Stage 3. Teachers View Muddy Points
The same 19 teachers from stage 1 returned to the lab for a second session approximately one week later, for one hour. Teachers first reviewed the video we chose from the one or two videos they produced during their first session. Then they were told how muddy points were generated for their lecture videos in each interface. Mudslide was referred to as the point-and-click interface, while the baseline was referred to as the free-response interface.

Teachers were given a quick tutorial of how to read the four components present in the teacher interfaces: the heatmap (Mudslide only), raw comments, histogram of frequent words, and Word Tree. Then, teachers saw (1) the muddy points generated and visualized with the baseline interface and (2) the muddy points generated and visualized in the Mudslide interfaces. The order of which interface each teacher saw first was counterbalanced across all teachers.

Teachers explored each interface themselves for approximately ten minutes while talking aloud to an experimenter taking notes. Their task during this time was to fill out a questionnaire about their impressions. After having seen both interfaces, teachers filled out an additional questionnaire asking for explicit comparisons between the two interfaces. As part of that survey, the experimenter reminded the teachers of their prediction about the muddiest point of their lecture videos (which they had made at the end of stage 1). They were then asked to reflect on the difference between what they had predicted and what they observed in students' muddy points. The lead author analyzed students' and teachers' free-text questionnaire responses for common themes using an iterative open coding approach; categories were not mutually exclusive.

## RESULTS

### Stage 1: Teachers Create Lecture Videos
The teachers made a total of 36 videos, which had a mean length of 3.6 minutes ($\sigma=2.4$), and 10.4 slides ($\sigma=5.3$). The 19 videos selected for stage 2 covered a range of topics, such as "Characteristics of Life," "Energy and Matter," "Six Traits of Writing," and "Resumes: How to Make Yourself Look Good."

### Stage 2: Students Generate Muddy Points
Nine of the 25 student participants indicated that their teachers at least occasionally assign videos to watch at home. Those nine students were then asked to describe what they do not like about watching lecture videos. Two students described the frustration of not being able to immediately ask a question. Five students complained that lecture videos could be too long or repetitive compared to reading a textbook. The remaining two students were frustrated by videos that were overwhelming, required pre-requisites they did not have, or used unfamiliar terms. When these nine students had questions triggered by an assigned lecture video, they relied on a combination of emailing their teacher (55%), using a search engine (55%), waiting for the next day's class (33%), or asking their friends (11%).

When reviewing the baseline and Mudslide student interfaces separately, students did not find either interface significantly more tedious or difficult to use, nor did they perceive a difference in the ease of remembering muddy points from the lecture. However, when asked which interface students preferred, 21 of the 25 students preferred the Mudslide student interface, and only four of the 25 students preferred the baseline interface.

*Baseline Student Interface*
Two main themes emerged from students' free responses describing their likes, dislikes, and wishes for the baseline interface. First, there is the freedom of the text box. 24% of students specifically mentioned how easy it was to review the video and simply type up their muddy points in the text box. This ease of use is perhaps especially true when muddy points are inherently non-spatial: *"It was easy to describe simple things such as voice volume or clarity of speech."* [S15] 36% of students liked that they *"could write down whatever [they] wanted"* [S10] to *"fully explain"* [S4] the muddiest point with *"a detailed reason."* [S22] An additional student [S14] appreciated that muddy points were not tied to a specific slide.

This freedom had costs, as well. When asked what they did not like about the baseline interface, 28% of the students specifically mentioned the difficulty or tedium of expressing the muddiest point only through text *"instead of just pointing to it."* [S3] The same student [S15] who extolled the simplicity of writing about *"simple things,"* like voice volume, observed that the text box alone is not suitable for more complex material: *"Advanced scenarios were very difficult to explain and say what about them was confusing."* When also considering what they wished for in this interface, 20% described a desire to anchor their comments to a particular slide or part of a slide, as is possible in the Mudslide interface.

24% of students specifically disliked not having thumbnails or wished for thumbnails of slides. Students were able to revisit the lecture video, but it *"was much easier to provide the muddiest moment when I had the visual cues to help remind me what they were."* [S4] As a final note, one student added, *"This method* [baseline UI] *of giving feedback was similar to the ones I've used in the past for lecture videos, so I was more accustomed to it, but I do not like it much."* [S2]

*Mudslide Student Interface*
When students wrote about their likes, dislikes, and wishes for the Mudslide student interface, the dominant two themes were ease and exactness. 44% of students described the system as easy to understand and use. 28% of students mentioned their appreciation for the exactness with which they could discuss the muddiest points: *"I really liked this method of entering the muddiest point. I could easily show the instructor the area where I was confused with pretty accurate precision"* [S4] and *"Very efficient, easy to show exactly what was confusing"* [S15]. Only 24% of students did not express appreciation for the system's ease or exactness in their response.

Requiring a spatial anchor for each muddy point affected the process of commenting on non-spatial aspects of the video. 16% of students described wanting to "point" to a moment in time or to what was being said over the slide, instead of the text or images on the slide. Without any summary representation of the audio as a timeline or transcript, students reported clicking somewhere on the slide containing the muddy voice-over, but could not be any more spatially specific than that. One student remarked that this was especially problematic when teachers spent a long time talking over a slide with little information on it.

The intentional constraint that every muddy point submitted had to be associated with a location on a slide caused frustration for the 8% of students who reported difficulty composing any muddy point. This task is not trivial: *"You had to come up with something very relevant and specific to the slides you were clicking on"* [S10]. One student, stuck, admitted to picking an arbitrary point.

While the simple click-and-describe muddy point submission process was clear and easy, it did not allow for students to indicate just how large or small the region on the slide they were "pointing" to was. 20% of students specifically complained that they could not add arrows or highlight large areas, even whole slides, at once.

*Muddy Points Generated*
To assess whether the student interface affected the type and quality of the muddy points generated by these students, two raters independently scored the muddy points generated by the 25 students. Since students were allowed to describe multiple muddy points in the baseline text box, each rater independently parsed how many muddy points the students' freeform textual response contained, and the category each parsed muddy point fell into. The categories chosen were Question, Description of Confusion or Complaint, Suggestion, Appreciation, and No Substance. Each parsed muddy point was given a single category. We defined comments of "no substance" as those comments containing no suggestions, questions, descriptions of confusion, complaints, or statements of appreciation. An example would be: "Nothing was confusing."

The raters also each marked muddy points as (1) either spatial (i.e., relating to an object on the slide) or non-spatial (i.e., relating to a non-spatial issue like volume), and (2) either specific (i.e., relating to a particular section or slide) or holistic (i.e., relating to the entire lecture). The same process was applied to the spatially anchored textual responses from the Mudslide student interface.

While one student's response in the baseline interface was lost, the 24 remaining freeform responses were parsed into 42 and 36 muddy points, respectively, by the independent raters. On the subset of 19 students' freeform responses that were parsed by both raters into the same number of muddy points, the categorical label agreement on that subset of 30 muddy points across raters was high (Scott's pi of 0.92) [16]. Of the 31 spatially anchored textual responses provided by the 25 students, the raters independently both parsed 32 muddy points (only one textual response had two muddy points contained within it). On those 32 parsed muddy points, the categorical label agreement was also high enough to be considered consistent (Scott's pi of 0.70) [16]. To merge their ratings, we took the sum of their respective category counts. Regardless of the interface used, students appropriately channeled the majority of muddy points into the intended categories of Questions and Confusions/Complaints. However, when using the Mudslide student interface, fewer students submitted muddy points of no substance (two-sided probability test, $p < 0.05$). Insubstantial muddy points like "Nothing was unclear" or "I was confused" are something that teachers mentioned as being uninformative.

The same raters coded muddy points as either spatial or non-spatial with a Cohen's κ [5] of 0.57 and 0.59 for the baseline and Mudslide interfaces, respectively. The raters coded muddy points as either specific or holistic with a Cohen's κ of 0.70 and 0.26 for the baseline and Mudslide interfaces, respectively. Agreement was fair or better, by a commonly cited scale [1], so we proceeded to test whether the interfaces had a significant effect on the specificity and spatial nature of muddy points. The total number of spatial comments (averaged across both raters) submitted with the Mudslide student interface was not significantly different than the analogous statistic for the baseline (one-tailed Fisher exact probability test). The same is true for specific comments. Across both interfaces, roughly a third of muddy points referred to a non-spatial aspect of the lecture video. A possibly overlapping third of muddy points across both interfaces refer to the lecture as a whole, rather than a specific part.

*Students' Reactions to the Mudslide Teacher Interface*
Twenty-one students had time left over within their session to explore the Mudslide teacher interface. Specifically, they explored the Mudslide teacher interface for the same lecture video for which they used the Mudslide student interface. On a 5-point Likert scale question (1-strongly disagree, 5-strongly agree) students felt that the Mudslide heatmap was easy to use (4.9, σ=0.4) and easy to understand (4.8, σ=0.6). Students disagreed with statements describing the heatmap as not useful (2.1, σ=1.1) and not interesting (2.0, σ=1.1). Students were comfortable with the idea of other students viewing their anonymous muddy points (4.3, σ=0.7) and strongly agreed with the statement "I like seeing other people's anonymous muddy points." (4.8, σ=0.5)

Students' free responses about the Mudslide heatmap flesh out these Likert scale ratings and contribute several additional themes. When asked what they learned while interacting with the heatmap, 52% of the 21 students specifically mentioned appreciating the opportunity to see the ideas of others. 29% of students appreciated the perspective they gained when seeing others' muddy points that were different from their own: *"I learned what others are confused about which helped with my personal analysis of potential energy"* [S25]. Finally, 33% expressed reassurance that, as one student wrote, *"people shared my thoughts on the muddiest section"* [S19].

When asked what they liked about the heatmap, the same theme of appreciating the access to everyone else's responses came through even more clearly: 86% of student responses mentioned this as a benefit. The remaining 14% use the space to describe the visual clarity or ease of use, e.g., *"Easy to see hotspots of problems."* [S15] When asked what they did not like about the heatmap, 43% listed nothing. Another third of students gave design requests, including the improvement of the slide's visibility beneath the heatmap and better filtering for or aggregation of similar muddy points.

**Stage 3: Teachers View Muddy Points**
In stage 3, teachers viewed the feedback generated by the 25 middle and high school students and the online crowd workers. The Mudslide teacher interface was populated with an average of 45.3 muddy cards (σ=16.7, min=24, max=94) from Turkers and 1.4 (σ=1.0, min=0, max=4) muddy cards from students per lecture video. The heatmaps generated for each lecture video were varied in appearance, depending on factors like the total number of muddy points (viewers had to specify at least one muddy point, but could specify as many as they liked if they found multiple parts of the lecture confusing), the number of slides in the lecture, and the information density on each slide. For example, the lecture video with the fewest slides introduced the concept of kinetic energy; the lecture contained only two slides, the second of which was quite information-dense, containing several equations. This lecture received 106 muddy points – three from three students, and 103 from 65 crowd workers. Shown in Figure 4, the vast majority, 93%, were on the second slide, and in particular were concentrated specifically over two of the eight equations on that slide, clearly indicating to the teacher which particular equations were not sufficiently explained in the video.

A different pattern of muddy points occurred in a lecture explaining how to set up a particular type of scientific experiment, which contained thirteen slides and received 60 muddy points (two from one student, and 58 from 39 crowd workers). On this lecture, the muddy points were more distributed across all slides, yet a clear plurality of points, 19, fell on a slide about the definitions of manipulated and responding variables (see Figure 1). In this case, the points were not clustered in one distinct sub-area within the slide, but the teacher could see that the text of these comments all reflected a similar theme relevant to the slide as a whole, e.g., *"Confusion over manipulated [sic] variable and responding variable."* As expressed during a session talk-aloud, the Mudslide design was in line with teachers' goals regarding feedback: *"If a student says, 'this is confusing' generally, I can't really help them. This way (with the heatmap), we can train them to be able to explain their confusion more clearly, which is something we try to teach."* [T1, paraphrased]

Over half of the teachers (53%) did not anticipate what the muddiest point of the lecture would be, indicating that muddy point feedback offers perspectives that are not easy

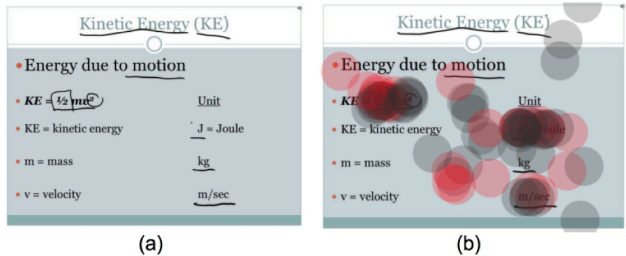

**Figure 4 (a)** Final slide in a video lecture on Kinetic Energy. **(b)** Students' and Turkers' muddy points, clustered around the KE equation, many of which are asking why there is a ½ or why velocity is squared, and around the symbol for Joule (J), asking for its definition. The teacher can now create targeted, data-driven clarifications.

for teachers to anticipate. Teachers' ability to foresee what might confuse students may depend on whether their online videos are watched by an audience comprised of unfamiliar viewers (as in our evaluation scenario) or a set of known students. Even without knowing the students and Turkers, teachers expressed excitement about the potential of rapid feedback *("LOVE the heat map!"* [T6]).

*Muddy Points as Formative Assessment for Teachers*
Teachers also saw muddy cards submitted by students written out as free-form responses. The baseline teacher interface displayed 30 ($\sigma=0$) muddy cards from Turkers and 1.3 ($\sigma=0.9$, min=0, max=3) muddy cards from students per lecture video. The number of Turker-produced muddy cards collected per lecture is less in the baseline condition but still comparable to typical classroom(s). After seeing both teacher interfaces, an older, partially retired teacher reflected on muddy points' utility for helping teachers improve their own skills: *"Since these days teachers have to collect portfolio information, the reflection they might be able to do on the effectiveness of a lecture could be greatly enhanced by this tool. By the same token, teachers might be intimidated by the feedback and reject or repress it or not want administrators to see it. Might be wonderful for peer reflection groups like CFG (Critical Friends Group)."* [T12] This teacher points out that tools like Mudslide could double as a formative assessment tool [19] for teachers, through which they could refine their lectures, in addition to directly helping students.

T12's musings proved accurate–during stage 3: teachers' emotional reactions to muddy points on their own lectures split teachers into two groups, one that interpreted the feedback constructively (~85%) and one that felt that their skills were not accurately reflected in the feedback from viewers who were not their actual students (~15%). The first group interpreted the feedback as a way to learn more about what worked and did not work in their lecture, and to learn about their teaching style in general. For example, *"What I thought was entertaining was, in fact, confusing. Analogies were too silly. The repetition of [the comment] 'This was a strange lecture' definitely makes me reflect on how I present information"* [T14]. Another teacher, faced with a hotspot of muddy points on one particular slide about dependent and independent variables within an example of an experiment, reflected that she would *"provide examples of identifying variables from other experiments"* [T2]. Another teacher, also looking at the heatmap, noted: *"I could see where my weaknesses were in the presentation, easily"* [T5]. However, the second, smaller group of teachers perceived the confusion as caused only by viewers' missing prerequisites. After the study, one teacher expressed concern that their ability to teach was being evaluated unfairly, as if the volume and character of the muddy points were directly related to their teaching ability.

After seeing both the baseline and Mudslide teacher interfaces, teachers reflected on how the muddy points affected their perception and design of their video lectures. On a 5-point Likert scale question (1-strongly disagree, 5-strongly agree), teachers expressed a strong (4.4, $\sigma=0.7$) desire to make changes to their lectures after seeing the muddy points. The muddy points changed their perceptions of the clarity of their lectures (4.2, $\sigma=0.9$). In spite of their presumably constrained time, teachers agreed with the statement, "I would make time to review this feedback from students during a typical day" (3.9, $\sigma=0.7$).

Teachers rated their agreement with statements explicitly comparing the baseline with Mudslide, e.g., "The feedback I got from the point-and-click responses was *more useful* than the feedback I got from students' free responses." The opposite statement, that the students' free responses were more useful than the point-and-click responses, was also included, to make the overall perceived bias of the survey neutral between the two interfaces. These opposing statements were not adjacent to each other, and were interspersed with other 5-point Likert scale questions about the interfaces. The same was done for statements explicitly comparing which interface "gave me a better sense of my students' confusion."

By the two-sided Wilcoxon rank sum test, Mudslide was rated significantly more useful (W = 262, $p < 0.05$) and gave teachers "a better sense of students' confusion" (W = 293, $p < 0.001$) than the baseline. These significance findings are backed up by teachers' questionnaire responses, such as, *"[The Mudslide UI] allowed me to see EXACTLY [emphasis theirs] where the kids were confused while [the baseline] method was more vague"* [T3]. However, teachers still noted some drawbacks; in the baseline interface, *"the students provided information about the pace and tone quality of the lecture that were not evident in the point and click feedback [Mudslide] web page"* [T12].

*Comparing Visualizations of Muddy Points*
We asked teachers to rank the four visualizations of muddy points available to them across the two interfaces (note that the heatmap was only available for muddy points collected with the Mudslide student interface). Each visualization was assigned a 4 for every time it was ranked 1st, a 3 for 2nd, etc., so that higher averages correspond to better rankings. The average rank given to the heatmap (3.4) and raw comment (3.3) visualizations of muddy points were both greater than the average rank of the Word Tree (1.9) and histogram (1.4) visualizations. A Friedman test revealed a significant effect ($p < 0.001$). Post-hoc two-sided Wilcoxon rank sum tests showed a significant difference between both the heatmap and the raw comments and the lower ranked Word Tree and histogram ($17.5<W<32.5$, $p < 0.001$).

The heatmap was repeatedly praised by teachers as a time-saver: *"I found that the heatmap was much better for a quick at a glance view, and I could literally zoom in on problem areas. If I had more time I'd prefer the comments in text form but I am pretty sure that doesn't happen very often..."* [T8] An average of 50 muddy points were generated per video lecture, so it was not prohibitive to skim every muddy point available, but it was not necessarily fast. Another teacher wrote, *"Reading through all of this takes a while. I think it may be helpful occasionally but it's not something I'd want all the time"* [T8].

When justifying their visualization rankings, 53% of teachers mentioned either the specificity of spatial annotations or the speed of visually identifying trouble spots as major advantages of the Mudslide teacher interface. Without the spatial organization offered in Mudslide, it was *"hard to just jump to a confusion point visually, you have to sort it out"* [T1]. On the other end of the utility spectrum, *"The word tree and histogram are interesting, but not that useful to me,"* another teacher [T2] wrote, echoing the sentiment of many others.

**DISCUSSION**

Student and teacher feedback from our study indicates that Muddy Card-inspired feedback can be a valuable addition to online video lectures. Taking advantage of the visual nature of the video medium by allowing students to anchor their muddy points spatially rather than merely typing a free-text response, seems like a particularly promising approach–this approach was preferred by students, resulting in fewer comments lacking substance, and enabled teachers to get a quick gestalt of a class's feedback via a heatmap representation.

Although Mudslide's visual point-and-click interface was suggestive of spatially-oriented feedback, many students nonetheless provided feedback on non-spatial and/or holistic aspects of the lecture. Providing visual representations of audio tracks (such as a graph of audio levels over time) or other meta-data about a lecture may be a valuable enhancement to our design. Although the simple circle-based representation of muddy points employed by Mudslide was effective, exploring ways to allow students to specify regions of varied shapes and sizes while preserving the simplicity of the specification process and the ease of creating a glance-able heatmap for teachers is a design challenge worth considering.

In our study, the students watching the videos were not members of the producing teachers' regular classes. In many "flipping" scenarios, students watch content produced by teachers they do not have an ongoing relationship with (e.g., a teacher may use a video by a well-known expert on a particular subject), so our study design is a reasonable facsimile of that scenario. However, in other flipping scenarios, a teacher may show her class self-produced videos–her knowledge of the students' backgrounds and prior contexts, and their knowledge of her teaching style may change the nature of the videos themselves and/or the type and interpretation of comments. The teachers' comments in the final stage of our study indicated that they found the feedback (from both the students and Turkers) to be similar to the types of feedback they get from their regular students, indicating that this may not be a large confound; however, investigating whether Mudslide-inspired feedback systems for familiar teacher-student sets necessitate different affordances than those for unfamiliar sets is an interesting avenue for future investigation.

Investigating whether Mudslide-like interactions would be a valuable feedback mechanism for MOOCs, whose viewership may be orders of magnitude larger than the classroom-flipping scenario we focused on, is also a valuable area for further study. Text summarization features, such as Word Trees and histograms, may be more valued by MOOC instructors than the teachers in our scenarios, as skimming individual comments becomes less feasible when thousands of students provide muddy points. Enabling filtering of the heatmap (perhaps allowing the teacher to remove points corresponding to certain confusion levels, or to students meeting certain criteria such as grades on prior class assignments, or perhaps based on tags that students could be encouraged to associate with their comments) may also become important to retaining its value as a tool for quick insights in a MOOC scenario.

Although traditional Muddy Cards are meant for the teacher's attention, students expressed interest in seeing others' muddy points. In our current design, a teacher could choose to send his URL for the Mudslide interface to the class if he wished to offer his students this experience. Considering how the muddy-point viewing interface might be redesigned specifically for this peer-oriented scenario (which may require higher standards for language blacklisting, privacy guarding, etc.) is another potential avenue of future study. This scenario might be particularly relevant for MOOCs, where peer-based interactions sometimes supplant teacher interactions for reasons of scalability [15].

## Assumptions and Limitations

We drew inspiration from literature and interviews about traditional lectures, but we designed Mudslide for short, slide-based, video lectures. Mudslide was not thoroughly tested on different scales of lecture length or number of muddy points. We are exploring Mudslide's flexibility.

## CONCLUSION

As teachers increasingly flip some or all of their classes by assigning students to view videos that they or other instructors have created as a homework activity [12], there is a need to capture signals about students' understanding of these videos. We identified an opportunity for interaction design by observing that the paper-based Muddy Card technique is well suited to adaptation to online video lectures. Our Mudslide prototype enhances the muddy card concept by taking advantage of the visual nature of video lectures to enable spatially anchored feedback; students preferred this to free-form feedback, and produced higher quality comments with this interface. The spatial nature of the feedback made glance-able confusion heatmaps possible, which teachers found valuable and more efficient to interpret than a list of free-form comments. Mudslide provides a simple, yet valuable, interface for enriching online video lectures by providing a way for students to efficiently and specifically express their confusion to instructors.

## ACKNOWLEDGMENTS

The authors would like to thank Anoop Gupta, Kurt Berglund, Sasa Junuzovic, and the Office Mix team for their feedback and support.

## REFERENCES

1. Altman, D.G. *Practical statistics for medical research.* CRC Press, 1990.
2. Angelo, T.A. and Cross, K.P. *Classroom assessment techniques.* 1993.
3. Breslow, L., Pritchard, D.E., DeBoer, J., Stump, G.S., Ho, A.D., and Seaton, D.T. Studying learning in the worldwide classroom: Research into edX's first MOOC. *Research & Practice in Assessment 8*, (2013), 13–25.
4. Brush, A.J.B., Bargeron, D., Grudin, J., Borning, A., and Gupta, A. Supporting Interaction Outside of Class: Anchored Discussions vs. Discussion Boards. *CSCL 2002*, International Society of the Learning Sciences (2002), 425–434.
5. Cohen, J. A Coefficient of Agreement for Nominal Scales. *Educational and Psychological Measurement 20*, 1 (1960), 37–46.
6. Craig, S., Graesser, A., Sullins, J., and Gholson, B. Affect and learning: an exploratory look into the role of affect in learning with AutoTutor. *Journal of Educational Media 29*, 3 (2004), 241–250.
7. Cross, A., Bayyapunedi, M., Ravindran, D., Cutrell, E., and Thies, W. VidWiki: Enabling the Crowd to Improve the Legibility of Online Educational Videos. *Proceedings of the 17th ACM Conference on Computer Supported Cooperative Work & Social Computing*, ACM (2014), 1167–1175.
8. D'Mello, S., Lehman, B., Pekrun, R., and Graesser, A. Confusion can be beneficial for learning. *Learning and Instruction 29*, (2014), 153–170.
9. Freeman, S., Eddy, S.L., McDonough, M., et al. Active learning increases student performance in science, engineering, and mathematics. *PNAS 111*, 23 (2014), 8410–8415.
10. Hall, S.R. Teaching By Questioning. *MIT Aero-Astro Annual*, 2003, 29–35.
11. Harwood, W.S. The one-minute paper: a communication tool for large lecture classes. *J. Chemical Education 73*, 3 (1996), 229.
12. Herreid, C.F. and Schiller, N.A. Case studies and the flipped classroom. *J. College Science Teaching 42*, 5 (2013), 62–66.
13. Kibler, J. Cognitive Disequilibrium. In S. Goldstein and J.A. Naglieri, eds., *Encyclopedia of Child Behavior and Development*. Springer US, 2011, 380–380.
14. Kim, J., Guo, P.J., Cai, C.J., Li, S.-W.D., Gajos, K.Z., and Miller, R.C. Data-Driven Interaction Techniques for Improving Navigation of Educational Videos. *UIST 2014*, .
15. Kulkarni, C., Wei, K.P., Le, H., et al. Peer and self assessment in massive online classes. *ACM Trans. on Computer-Human Interaction 20*, 6 (2013), 33.
16. Lombard, M., Snyder-Duch, J., and Bracken, C.C. Practical resources for assessing and reporting intercoder reliability in content analysis research projects. *Retrieved April 19*, (2004), 2004.
17. Luther, K., Pavel, A., Wu, W., et al. CrowdCrit: crowdsourcing and aggregating visual design critique. *CSCW 2014*, ACM (2014), 21–24.
18. Mosteller, F. The 'Muddiest Point in the Lecture'as a feedback device. *On Teaching and Learning: J. Harvard-Danforth Center 3*, (1989), 10–21.
19. Sadler, D.R. Formative assessment and the design of instructional systems. *Instructional science 18*, 2 (1989), 119–144.
20. Surowiecki, J. *The wisdom of crowds.* Random House LLC, 2005.
21. Wattenberg, M. and Viegas, F.B. The Word Tree, an Interactive Visual Concordance. *IEEE Transactions on Visualization and Computer Graphics 14*, 6 (2008), 1221–1228.
22. Wilson, R.C. Improving faculty teaching: Effective use of student evaluations and consultants. *J. Higher Education*, (1986), 196–211.
23. Xu, A., Huang, S.-W., and Bailey, B. Voyant: generating structured feedback on visual designs using a crowd of non-experts. *CSCW 2014*, ACM (2014), 1433–1444.
24. Zyto, S., Karger, D., Ackerman, M., and Mahajan, S. Successful classroom deployment of a social document annotation system. *CHI 2012*, ACM (2012), 1883–1892.